\documentclass[11pt,a4paper]{article}
\usepackage[utf8]{inputenc}
\usepackage{geometry}
\usepackage{graphicx}
\usepackage{hyperref}
\usepackage{booktabs}
\usepackage{xr}
\usepackage{tcolorbox}
\tcbuselibrary{listings} % <--- Add this line right below it!
\usepackage{listings}

\newtcblisting{promptbox}{
  arc=2mm,                  % Rounded corners
  boxrule=0.5pt,            % Border thickness
  colback=gray!5,           % Background color
  colframe=gray!60,         % Border color
  listing only,
  listing options={
    basicstyle=\small\ttfamily,
    breaklines=true,        % Wrap long lines automatically
    columns=fullflexible
  }
}

\title{Large-scale dataset of automatically classified rhetorical sections in scientific papers}
\author{
    Daniel Verdi$^{1,2,3}$, Jacob Aarup Dalsgaard$^{2,3}$, Roberta Sinatra$^{2,3,4,5}$ \\
    \small{$^{1}$Graduate School of Education, Stanford University, United States} \\
    \small{$^{2}$Center for Social Data Science (SODAS), University of Copenhagen, Denmark} \\
    \small{$^{3}$Networks, Data, and Society (NERDS), IT University of Copenhagen, Denmark} \\
    \small{$^{4}$Pioneer Centre for Artificial Intelligence (P1), Denmark} \\
    \small{$^{5}$Complexity Science Hub, Austria} \\
    \small{$^{*}$Corresponding authors: verdi@stanford.edu, jad@sodas.ku.dk, robertasinatra@sodas.ku.dk}
}
\date{}

\begin{document}

\maketitle

\begin{abstract}
Scientific papers follow rhetorical structures that organize content into sections such as Introduction, Methods, Results, and Discussion. Automatically identifying these sections at scale enables granular analysis of scientific writing patterns. We present a dataset of section-level annotations for millions of scientific papers from the Semantic Scholar Open Research Corpus (S2ORC). Using a rule-based classification algorithm, we identified and labeled major sections across 15.6 million papers after quality filtering. The dataset covers primarily STEM disciplines, with strong representation in medicine and biology. We provide comprehensive human and LLM-based validation showing that classifier agreement with human annotators is on par with human inter-annotator agreement. This dataset enables large-scale computational studies of scientific discourse and writing patterns.
%including a detailed study of AI tools adoption in different parts of the research communication process.
\end{abstract}

\section{Background \& Summary}

Understanding how scientists communicate their findings requires analyzing the rhetorical structure of scientific papers. Researchers often organize their work into standardized sections that serve distinct communicative purposes, such as introductions framing the research problem, methods describing experimental procedures, and results presenting findings, many times following the IMRaD format (Introduction, Methods, Results, and Discussion) \cite{imrad2004, imrad2024}. Automatically identifying these sections at scale enables new forms of analysis that would be impractical with manual annotation. However, creating such section-level annotations for large corpora presents significant technical challenges due to the variability in section headers across disciplines, publishers, and document structures.

Several researchers have previously addressed this challenge, though with important limitations. Shahid and Afzal \cite{shahid2018section} used keyword matching combined with logical section order, such as assuming the introduction appears first. Treeratpituk et al. \cite{treeratpituk2010seerlab} and Tuarob et al. \cite{tuarob2015hybrid} extended the keyword approach by employing regular expressions to capture broader variations of section-related terms and phrases. Nguyen and Kan \cite{nguyen2007} developed a Maximum Entropy classifier using four features: section number, relative position within the document, previous section header, and current section header. More recently, Rahman and Finin \cite{rahman2019_unfolding} applied a word-based convolutional neural network model, converting input texts into multi-label one-hot vectors passed through an embedding layer.

These prior approaches share several constraints that limit their applicability to large-scale corpus analysis. Most were evaluated on relatively small datasets \cite{shahid2018section, treeratpituk2010seerlab, tuarob2015hybrid}, often containing only hundreds of papers, making it unclear whether they generalize to the diversity encountered in million-scale corpora. Additionally, none of these research efforts made their code publicly available, preventing replication or extension of their methods.

We address this by developing a classification algorithm adapted to the scale and characteristics of the Semantic Scholar Open Research Corpus (S2ORC) \cite{lo2020s2orc}, which contains the full text of 15.6 million open-access scientific papers. Although we apply the method to S2ORC due to data availability, the approach is applicable to any scholarly corpus that provides structured full-text documents parsed through GROBID or similar PDF-to-XML parsing pipelines.

\begin{figure}[h]
    \centering
    \includegraphics[width=0.85\textwidth]{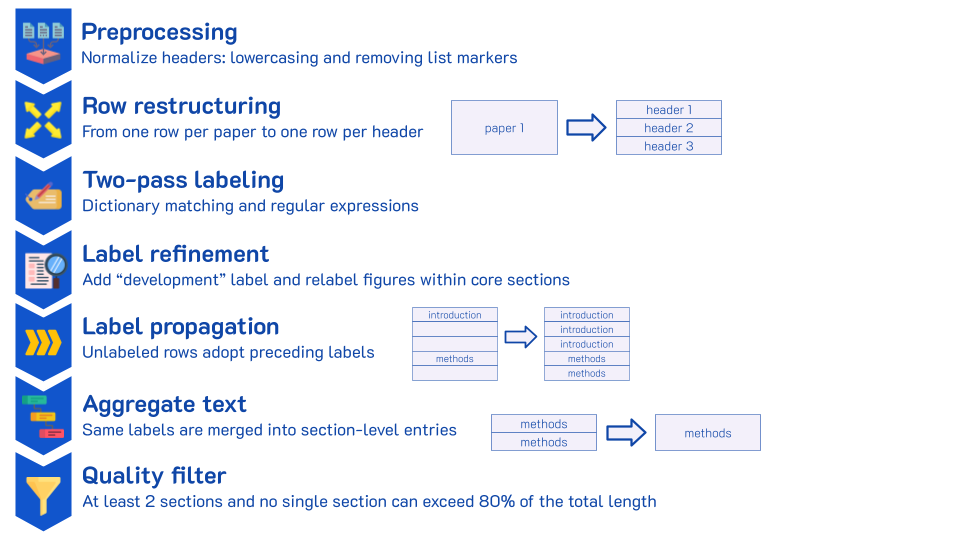}
    \caption{Simplified diagram of the main steps of the algorithm. Icons from \href{https://www.flaticon.com/}{Flaticon.com}}
    \label{fig:diagram}
\end{figure}

Our approach (summarized in Fig.~\ref{fig:diagram}) builds on prior work while addressing the specific constraints of large-scale processing. We use rule-based methods combining exact matching and regular expressions, informed by the pattern-matching approaches of earlier studies, but designed for computational efficiency and transparency at scale. The algorithm identifies major section boundaries through two-pass labeling, followed by label propagation and filtering steps to ensure quality.

After filtering, our dataset contains section annotations for 13.5 million papers, representing 86 percent of the initially classified documents. This scale represents two orders of magnitude more annotated papers than prior datasets, enabling statistical analyses and machine learning applications that require large training or evaluation sets. The dataset focuses on empirical papers following IMRaD-like structures, with labels for introduction, methods, results, discussion, conclusion, and supporting sections.

We validated our approach through human annotation of 100 papers and Large Language Model (LLM)-based annotation of 1,000 papers, achieving Krippendorff's Alpha agreement scores of 0.61 between human annotators and the algorithm on cleaned labels. The resulting dataset mostly represents science, technology, engineering, and mathematics (STEM) disciplines, comprising approximately 75 percent of all papers.

This resource enables researchers to conduct a range of Science of Science \cite{fortunato2018science} studies, such as large-scale computational studies of scientific discourse, examination of discipline-specific writing conventions, and tracking changes in scientific communication over time at the section level. Indeed, the combination of scale, validation, and public availability addresses key gaps in existing resources for analyzing the rhetorical structure of scientific literature.

\section{Methods}

\subsection{Data Source}

We used the Semantic Scholar Open Research Corpus (S2ORC) \cite{lo2020s2orc} as our primary data source. S2ORC aggregates full-text content from hundreds of academic publishers and digital archives, containing 15,679,781 scientific papers with parsed text and metadata. The data was retrieved using the Semantic Scholar Datasets API \url{https://api.semanticscholar.org/api-docs/datasets} using release 2026-05-26. The corpus processing pipeline converts PDF files (and LaTeX source when available) into structured XML format using GROBID, a machine learning library for extracting and parsing scholarly documents \cite{grobid}. This structured format includes annotations for citations, section headers, and text spans indicating their positions within documents.

The S2ORC dataset is heavily concentrated in STEM disciplines (natural sciences, computer science, engineering, math). We computed these statistics from the field-of-study annotations provided in S2ORC, which are derived from Semantic Scholar’s S2FOS classification model \cite{s2_fos}. From that, we observed that 89\% of papers are assigned to STEM-related fields. Medicine forms by far the largest proportion, representing 25.6\% of all papers, followed by Environmental Sciences (10.4\%), Biology (9.0\%), and Engineering (8.3\%). Additionally, although S2ORC is intended to be English-language, we observe that approximately 8–10\% of documents are non-English text.

While we use S2ORC as our primary data source, the proposed approach is applicable to any scholarly corpus that provides structured full-text documents generated through GROBID or a similar PDF-to-XML parsing pipeline. Examples of corpora that provide comparable full-text representations include CORE \cite{core} and part of the OpenAlex database \cite{openalex}. Consequently, the approach is not specific to S2ORC and can be applied to other large-scale scholarly collections with similar structured document formats and an expected IMRaD structure.

\subsection{Data Preparation}

We processed S2ORC data files in Parquet format through several preparation steps. First, we preprocessed GROBID-extracted headers by converting text to lowercase and removing list markers from the beginning of strings. This normalization step accounts for variations in header formatting across different publishers and document processing artifacts (e.g. "I. Introduction" and "1. INTRODUCTION" both become "introduction").

We then restructured the data from a paper-level representation (15 million rows) to a header-level representation (over 200 million rows), where each row corresponds to a single header within a paper. To manage memory constraints while processing this expanded dataset, we partitioned the data into smaller batches for parallel processing.

For each header row, we extracted the associated text content spanning from that header to the start of the next header. This segmentation approach defines boundaries based on the structural markers already identified by the S2ORC parsing pipeline. The extracted text for each section was stored separately to optimize memory usage during classification.

\subsection{Two-Pass Labeling Strategy}

Our classification approach uses rule-based methods in two passes. The first pass employs exact string matching against a dictionary of common, generic section titles, including terms like ``results," ``conclusion," ``methods," and ``introduction." This conservative exact-matching approach captures headers using standard terminology without variation. The list of words was obtained from manual inspection of the most common words in the GROBID-extracted headers that could be clearly classified into standardized sections.

The second pass applies regular expression patterns to identify broader variations of section headers that maintain semantic equivalence to standard labels. The regular expressions were developed through exploratory analysis of common header patterns in the corpus. For example, patterns capture variations like ``study design", ``study description", and ``proposed implementation" as instances of methods sections. The challenge in designing these patterns was to capture sufficient variation while avoiding overly broad matches that would lead to misclassification. Terms that are too general (such as ``problem") were excluded to prevent false positives.

Our approach is based on rule-based methods rather than more complex machine learning methods or large language models for several reasons. First, rule-based classification provides transparency and controllability, allowing direct understanding of why specific labels were assigned. Second, manual analysis of unlabeled rows revealed that difficult cases not captured by the rule-based approach depend heavily on contextual information from surrounding text and previous headers. This suggests that our method may be approaching the performance ceiling achievable using only header text extracted by GROBID. Improving classification accuracy would require providing substantially more contextual information, such as full section text, citation patterns, or document-level features, thereby increasing both computational costs and annotation complexity while potentially still yielding ambiguous classifications for papers with non-standard structures. Third, the computational efficiency and cost advantages of rule-based methods are substantial at this scale. Processing 15 million papers with our rule-based pipeline required minimal computational resources and completed within hours on standard hardware, whereas machine learning approaches would require extensive feature engineering and training time, and LLM-based classification would incur prohibitive API or GPU costs.
%(in the order of thousands of US dollars).
The rule-based approach thus provides an optimal balance between classification accuracy, transparency, scalability, and resource efficiency for corpus-scale annotation.

Our classification schema focuses on empirical papers with IMRaD-like structures. The adopted labels include: introduction, methods, results, discussion, conclusion, figure\_table (for figure and table captions), ending (for references, appendices, and similar back matter), and other (for sections like conflicts of interest and funding statements).

\subsection{Label Refinement}

After the initial two-pass labeling, many header rows remained unlabeled, including both subsection headers within major sections and major section headers that did not match our dictionaries or patterns. We applied label propagation to assign labels to these remaining rows, assuming that unlabeled rows between two labeled rows belong to the section of the preceding label. For example, if row 4 is labeled ``methods" and row 16 is labeled ``results," rows 5 through 15 are assigned the ``methods" label.

This propagation strategy required handling two types of special cases. First, non-core headers such as figure captions, table labels, and ethical statements appear both within main narrative sections and as separate end matter. When these elements appear within the main body (for example, a figure caption in the results section), we relabel them with the major section label (in the example, ``results"). When they appear in trailing end matter blocks, they retain their original labels ("figure\_table" in this example).

%%Second, label propagation assumes that all major sections have been identified by the initial labeling passes. To account for unlabeled major sections, particularly in more argumentative papers that do not follow strict IMRaD structure, we created a ``development" section label. This label identifies major sections following the introduction that were not captured by our standard patterns, using numerals (for example, ``2. Historical Background") as indicators of top-level section headers. While only about 30 percent of papers use consistent numeral-based header formatting, this approach helps identify where introduction sections end, particularly when the subsequent section uses discipline-specific terminology not captured in our dictionaries.

To account for papers whose organizational structure does not follow standard IMRaD conventions, we introduce a \textit{development} label. This label is assigned to major top-level sections that are not matched by any of our section dictionaries but are identified through structural cues, primarily numbered section headers (e.g., ``2. Historical Background''). The label therefore serves as a generic category for substantive sections that continue the main argument or exposition of the paper without corresponding to a standard rhetorical function such as methods, results, or discussion. In practice, the \textit{development} label plays an additional role by helping identify the boundary of the introduction. Many papers, particularly in the humanities and social sciences, transition from the introduction into sections with discipline-specific titles that cannot be reliably classified using rule-based dictionaries alone. By identifying these major unlabeled sections, the \textit{development} category allows the algorithm to recognize where the introduction ends while preserving the distinction between introductory material and the remainder of the paper. The label should therefore be interpreted primarily as an indicator of a major non-standard section rather than a specific rhetorical category.

\subsection{Text Aggregation and Filtering}

After label assignment and refinement, we aggregated text from consecutive rows with identical labels within each paper. This consolidation transforms multiple header-level rows into single section-level entries, producing the final section-based representation of each document.

%We then applied filtering criteria to retain only papers meeting quality thresholds. Papers must have at least two identified sections to pass filtering, which removes documents where the classifier identified only a single section (typically non-empirical papers lacking the expected rhetorical structure). Additionally, no single section can exceed 80 percent of the adjusted paper length, where adjusted length excludes trailing sections like references. This threshold filters papers where label propagation likely failed, resulting in most of the paper being incorrectly labeled as an introduction.

We then applied filtering criteria to retain only papers meeting baseline quality thresholds. First, papers must have at least two identified sections to pass filtering. This constraint removes documents where the classifier identified only a single section, which typically indicates non-empirical papers lacking the expected rhetorical structure. In our dataset, the median number of sections is four; requiring at least two sections retains 95.5\% of the corpus, whereas increasing this minimum to three would remove an additional 12\% of the documents. Additionally, we impose a threshold on the maximum proportion of the paper that a single section can occupy (measured against an adjusted length that excludes trailing sections like references). This threshold is designed to filter out papers where label propagation likely failed, which frequently results in the bulk of the paper being incorrectly labeled as an introduction.

To avoid setting a completely arbitrary maximum-length threshold, we systematically evaluated outlier detection strategies. Methods assuming normality, such as standard or MAD-based robust z-scores, proved unsuitable because the distribution of maximum section lengths is highly skewed, exhibiting a spike at 100\% (with a median of 49.9\%, a 75th percentile of 66.5\%, and a 90th percentile of 85.6\%). We also discarded strict percentile cutoffs, as they discard a fixed fraction of the data regardless of underlying classifier quality. Ultimately, we compared variations of the Interquartile Range (IQR) rule against the validation subset. As expected, retaining more documents tends to introduce lower-quality papers. For instance, a stricter IQR configuration ($k=0.25$, where $k$ is used to calculate papers to keep below $Q3 + k \times IQR$, yielding a threshold of 73.2\%) improves validation metrics (Macro F1 of 0.657, Weighted F1 of 0.741) but drops retention to 81.2\%. While we interpret the exact F1 curves cautiously due to the relatively small size of the validation samples, we ultimately adopted an IQR multiplier of $k=0.50$ as our final filtering criterion. This configuration provided the most stable and systematic balance between dataset size and document quality, retaining 81.2\% of the papers while preserving strong classification performance (Macro F1 of 0.657, Weighted F1 of 0.7411).

These filtering steps reduced the dataset from 15,679,781 initially classified papers to 13,530,774 papers in the final dataset, representing 86 percent retention. The analysis by field in SI section 2 indicates that the filtering criteria strongly favor disciplines utilizing traditional IMRaD structures. Fields like Computer Science and Engineering are retained at substantially higher rates than non-traditional fields like History and Philosophy, suggesting that the filter effectively isolates and removes papers lacking standard section headers.

\section{Data Records}

The dataset is organized into a single file containing the corresponding section labels for every header in each paper. Each record includes the paper identifier from S2ORC, the header-section label, and metadata indicating the header's position within the document. We do not include the full text content of each header in our dataset, as this would duplicate content already available in S2ORC. Users can retrieve the corresponding text by matching our paper identifiers and header position metadata to the original S2ORC corpus, ensuring compatibility while minimizing redundancy. Table~\ref{tab:section_classified_fields} describes the data fields of the full classified dataset used to represent header-section boundaries and their associated labels. Table \ref{tab:section_counts} reports the number of papers with each of the major sections in the dataset.

\begin{table}[h!]
\centering
\caption{Field descriptions for the full classified dataset.}
\setlength{\tabcolsep}{4pt}
\begin{tabular}{@{}p{0.20\textwidth} p{0.10\textwidth} p{0.20\textwidth} p{0.45\textwidth}@{}}
\toprule
\textbf{Column} & \textbf{Type} & \textbf{\# of Non-empty Values} & \textbf{Description} \\
\midrule
\texttt{corpusid} & integer & 313,975,048 & Semantic Scholar Corpus ID identifying the publication. \\
\texttt{start} & integer & 313,975,048 & Character offset marking the beginning of the section. \\
\texttt{end\_header} & integer & 313,975,048 & Character offset marking the end of the section header. \\
\texttt{end\_section} & integer & 313,975,048 & Character offset marking the end of the section. \\
\texttt{sec\_label} & string & 296,839,920 & Canonical section label assigned to the section. \\
\texttt{sec\_alt\_label} & string & 2,027,585 & Alternative section label extracted when section is composite. \\
\bottomrule
\end{tabular}
\label{tab:section_classified_fields}
\end{table}

\begin{table}[ht]
\centering
\caption{Number of papers by section label.}
\label{tab:section_counts}
\begin{tabular}{lr}
\toprule
\textbf{Section} & \textbf{Papers} \\
\midrule
Introduction & 11,723,065 \\
Methods      & 9,530,120 \\
Results      & 8,428,993 \\
Discussion   & 8,985,561 \\
Conclusion   & 8,738,357 \\
Development  & 3,116,908 \\
\bottomrule
\end{tabular}
\end{table}

Approximately 50 percent of papers contain three core empirical sections (introduction, methods, and one of results or discussion or conclusion), reflecting the dataset's focus on empirical research following IMRaD-related structures.

The data is available through a public repository (\url{https://zenodo.org/records/20814302}). The repository includes the complete header-level dataset in Parquet format, an aggregate metadata table, documentation of the label schema, and code for reproducing the classification pipeline. Table~\ref{tab:section_aggregated_fields} summarizes the data fields of the aggregated dataset.

\begin{table}[h!]
\centering
\caption{Field descriptions for the aggregated dataset.}
\setlength{\tabcolsep}{4pt}
\begin{tabular}{@{}p{0.20\textwidth} p{0.10\textwidth} p{0.20\textwidth} p{0.45\textwidth}@{}}
\toprule
\textbf{Column} & \textbf{Type} & \textbf{\# of Non-empty Values} & \textbf{Description} \\
\midrule
\texttt{corpusid} & integer & 69,670,197 & Semantic Scholar Corpus ID identifying the publication. \\
\texttt{sec\_label} & string & 66,997,994 & Canonical section label assigned to the text segment. \\
\texttt{paper\_len} & integer & 69,670,197 & Total length of the paper in characters. \\
\texttt{section\_length} & integer & 69,670,197 & Length of the section in characters. \\
\texttt{adj\_paper\_len} & integer & 69,670,197 & Paper length after excluding non-content text. \\
\texttt{perc\_of\_new\_total} & float & 69,670,197 & Section length expressed as a percentage of the adjusted paper length. \\
\texttt{s2fieldsofstudy} & string & 59,557,497 & Semantic Scholar field(s) of study associated with the publication. \\
\bottomrule
\end{tabular}
\label{tab:section_aggregated_fields}
\end{table}

\section{Technical Validation}

We conducted a technical validation using both human and LLM-based annotations. For the human validation, we randomly sampled 100 papers from the pool of 1,000 papers used for the LLM evaluation. We chose to sample complete papers rather than individual header rows because difficult classification cases often depend on the document context, including surrounding text and the sequence of previous and following headers.

\subsection{Human Annotation}

\begin{figure}[h]
    \centering
    \includegraphics[width=0.85\textwidth]{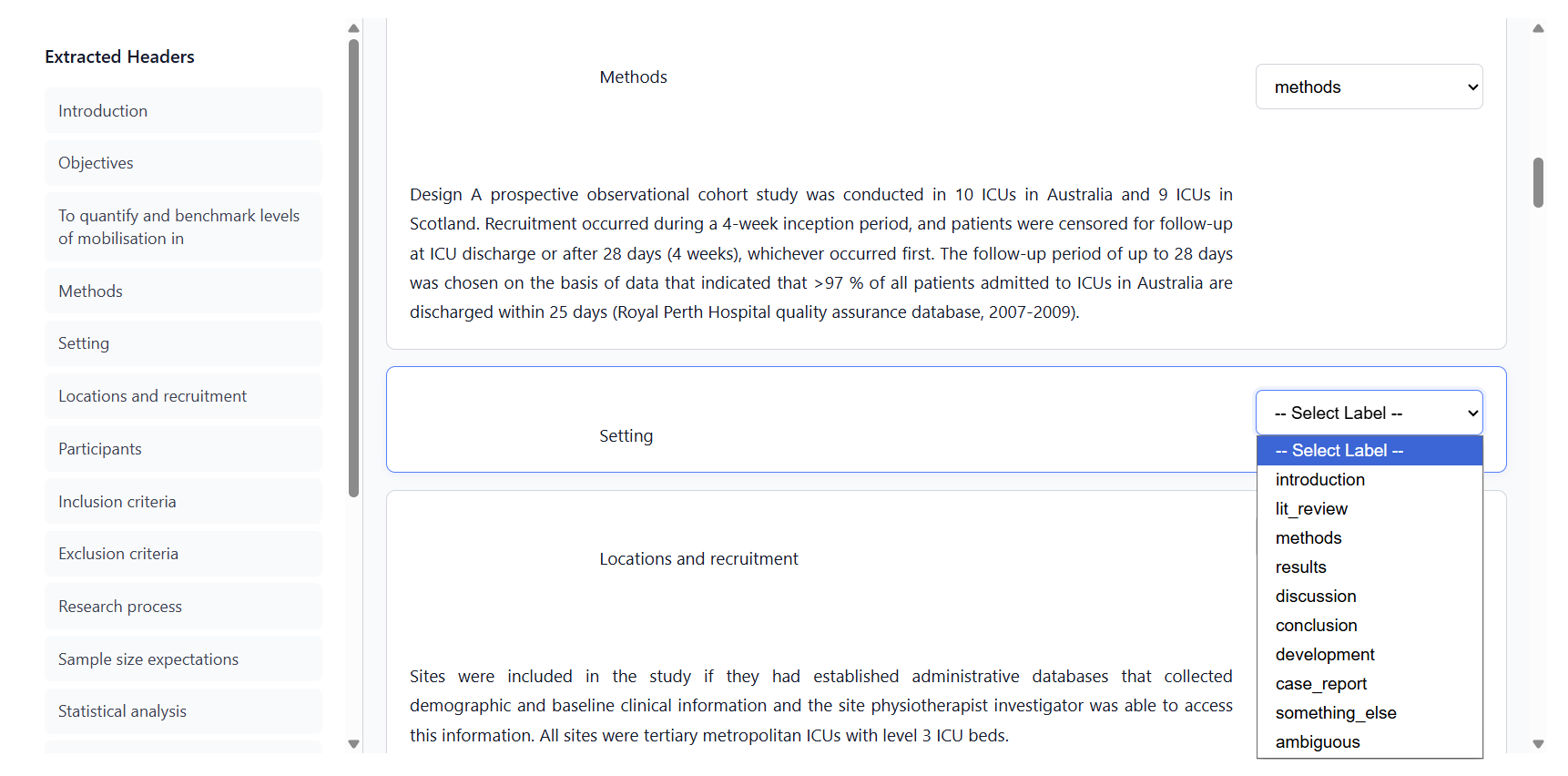}
    \caption{Annotation Interface showing the GROBID-extracted headers on the left, and text and selection boxes to the right.}
    \label{fig:interface}
\end{figure}

We recruited participants via the Prolific platform, targeting individuals with a minimum of a Master’s degree and English as their first language. These criteria were applied using Prolific's pre-screening filters and subsequently verified through self-reporting within our task interface; participants whose self-reported data did not match the required criteria were excluded. Annotators were compensated in accordance with the platform’s ethical wage guidelines. Figure \ref{fig:interface} shows the annotation interface.

To ensure data quality, candidates were required to pass a screening phase that involved practicing annotation on a control paper. This paper was identical for all participants and selected because its content could be unambiguously labeled according to our schema. Participants who failed to achieve at least 70\% accuracy on this practice task were granted one additional attempt; those who failed a second time were excluded from the study.

In total, 100 randomly sampled papers were labeled, with each paper receiving annotations from at least three unique participants. Each of the 32 final annotators labeled 10 papers. Two papers from the random sample were not in English and are not included in the analyses below.

\subsubsection{Participant Demographics}

According to the demographic data provided by Prolific, the participant pool had a mean age of 36.47 years ($SD = 10.73$, range: 19–62). All participants held advanced degrees, with 68.75\% (n=22) holding a Master’s degree and 31.25\% (n=10) holding a Doctorate. The sample was 65.62\% female and 34.38\% male. In terms of ethnicity, 65.62\% identified as White, 18.75\% as Black, and 15.62\% as Mixed. Participants were primarily born and resident in the United Kingdom (62.50\% born and 78.12\% residing) and the United States (21.88\% born and residing).

\subsubsection{Inter-annotator agreement and Evaluation Metrics}

The inter-annotator agreement yielded a Krippendorff’s Alpha of $\alpha = 0.61$, considered a low or moderate agreement. Rather than indicating poor data quality, we argue that this score serves as empirical evidence of the inherent complexity and subjective nature of rhetorical classification in scientific discourse. Even for highly educated native speakers, some of the section boundaries were ambiguous, or the titles likely used non-standard organizational schemes that resisted clear-cut categorization. 

Despite this complexity, the distribution of human labels across the 1,213 unique annotated items demonstrates a considerably high degree of functional consensus. While we observed unanimous agreement for 698 items (57.5\%), a majority agreement ($> 50\%$, i.e., in our case, meaning at least 2 out of 3 annotators assigning the same label) was present for 1,074 items (88.5\%).

\subsection{LLM Annotation}

For the LLM-based validation, we used the GPT-5-mini model via the OpenAI Batch API. We randomly sampled 1,000 papers for annotation.

To ensure the model had sufficient context for classification, each request included the target header along with a window of up to five preceding and five following headers. For headers at the beginning or end of a document, this window was dynamically truncated to include only available adjacent headers. Additionally, each prompt included the first 300 tokens of the section's content, tokenized using the o200k\_base encoding. We employed a one-shot prompting strategy, providing the model with a single labeled example and the exact same ten-category definitions provided to human annotators. The full prompt is included in the SI section 1.2.

We used OpenAI's Structured Outputs feature with a strict JSON schema requiring two specific fields: \textit{label} (the predefined section types) and \textit{confidence} (an enumerated string: \textit{high, medium, low}). By setting "strict: True" in the API configuration, we guaranteed adherence to this format.

The annotation was performed in three passes, and we used a majority-vote label when labels from different passes differed. The total cost for annotating the sampled headers was 10.8 USD. The LLM Inter-Run Agreement (Krippendorff's Alpha) was 0.9261.

\subsection{Evaluation Metrics}

Given the inherent difficulty of the classification task, we move away from reporting a singular ``accuracy" metric and instead evaluate our classifier using two approaches:

\begin{enumerate}
    \item \textbf{Treating human or LLM annotations as ``gold standard":} We compare the classifier’s predictions against a ``gold standard" dataset established by taking the majority vote of the three independent human or LLM annotators for each section. This measures the model's ability to align with the collective consensus (or the dominant human interpretation). We calculated precision, recall, and F1 scores for each section label.

    The decision to use majority-voting labels intends to retain rows with agreement, ensuring that evaluation metrics reflect performance against more clearly correct labels rather than very ambiguous cases.
    
   \item \textbf{Inter-rater Agreement:} We also measure agreement between all three annotators (humans, LLMs, and our rule-based classifier) without designating a ``ground truth." This treats the classifier as an additional independent annotator, which is plausible considering the already established difficulty of the task. For this approach, we computed Krippendorff's Alpha between annotation sources. By adopting this approach, we establish the human agreement level ($\alpha = 0.61$) as the functional upper bound for the task. We contend that a classifier achieving agreement levels comparable to this human baseline demonstrates a successful capture of rhetorical patterns at a level of consistency equivalent to expert human performance.
    
\end{enumerate}

% Additionally, we conducted evaluations both with all original labels (as-is evaluation) and after relabeling and merging certain categories. The relabeling addressed labels that were created during annotation but not available to the classifier (case\_report, something\_else, and ambiguous) and merged the development label with other categories since it was created primarily as an algorithmic aid rather than a substantive section type.

\section{Validation Results}

\subsection{Inter-Rater Agreement Scores}

As shown in Table \ref{tab:agreement_alpha}, our rule-based approach achieves agreement levels with human annotators ($\alpha = 0.5848$) that are remarkably comparable to the agreement between independent human experts ($\alpha = 0.6092$) and between humans and the LLM ($\alpha = 0.6293$).

\begin{table}[h]
\centering
\caption{Krippendorff's Alpha Agreement Scores} % The last two rows are filtered for classifier-supported labels
\label{tab:agreement_alpha}
\begin{tabular}{lcc}
\hline
\textbf{Comparison} & \textbf{Alpha ($\alpha$)} & \textbf{Items ($n$)} \\ \hline
Human vs. Human & 0.6092 & 1,213 \\
Human vs. LLM & 0.6293 & 1,126 \\
Human vs. Classifier & 0.5848 & 1,204 \\
LLM vs. Classifier & 0.5514 & 1,126 \\ \hline
\end{tabular}
\end{table}

\subsection{Classification Performance and Rhetorical Boundaries}

\begin{figure}[h]
    \centering
    \includegraphics[width=0.85\textwidth]{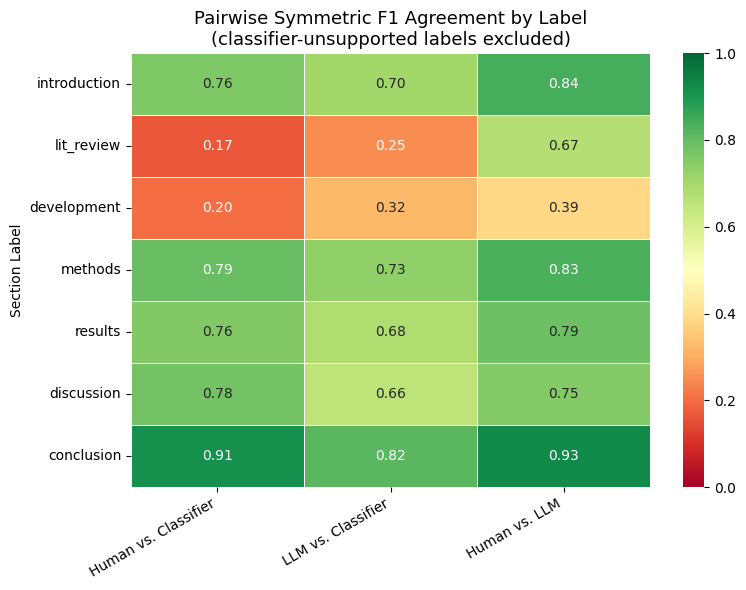}
    \caption{Pairwise Symmetric F1 Agreement by Section Label. Higher scores indicate stronger alignment between the two rater types for a specific category.}
    \label{fig:f1_heatmap}
\end{figure}

Beyond global agreement scores, we evaluate the classifier's performance across individual rhetorical labels by treating the majority-vote human annotations as the ground truth. Figure \ref{fig:f1_heatmap} illustrates the F1 agreement across all three rater types.

The results demonstrate that the rule-based algorithm achieves performance equivalent to the LLM within standard IMRaD sections. For \textit{methods} ($F1=0.79$) and \textit{conclusion} ($F1=0.91$), the algorithm aligns closely with human experts, suggesting that these sections are consistently identified through standardized terminology. This parity is significant as it confirms that, for core empirical reporting, our computationally efficient rule-based approach captures rhetorical patterns as effectively as more complex models.

In contrast, performance decreases for the \textit{lit\_review} and \textit{development} categories across all raters. The rule-based classifier's lower performance in these areas (e.g., $F1=0.17$ for \textit{lit\_review}) is primarily due to the non-standard, descriptive titles often given to these sections, which lack the explicit linguistic markers our dictionary-based approach relies on. While the LLM shows higher sensitivity to these categories, it notably also struggles with the \textit{development} label, mirroring the difficulty even human experts face in distinguishing these sections from the broader introduction.

\subsection{Precision and Recall Analysis}

To further decompose the classifier's performance, we analyze the precision and recall scores against the human majority-vote gold standard, as shown in Table \ref{tab:pr_metrics}. This granular view allows us to distinguish between categories where the classifier is conservative (high precision) versus those where it is broad (high recall).

\begin{table}[h]
\centering
\caption{Precision and Recall Metrics for Rule-Based Classifier (vs. Human Gold Standard)}
\label{tab:pr_metrics}
\begin{tabular}{lcc}
\hline
\textbf{Section Label} & \textbf{Precision} & \textbf{Recall} \\ \hline
Introduction & 0.67 & 0.89 \\
Lit\_review & 0.69 & 0.09 \\
Development & 0.12 & 0.79 \\
Methods & 0.86 & 0.73 \\
Results & 0.86 & 0.67 \\
Discussion & 0.74 & 0.82 \\
Conclusion & 0.86 & 0.97 \\ \hline
\end{tabular}
\end{table}

The \textit{Introduction} section exhibits a high recall ($0.89$), which aligns with our label propagation strategy: by defaulting unlabeled early sections to the introduction, we ensure that the rhetorical start of the paper is captured, even if some subsections (like literature reviews) are subsumed within it. Conversely, standard IMRaD sections like \textit{Methods} and \textit{Results} show a high precision ($0.86$), confirming that when the classifier matches a string, it is highly likely to be correct.

The trade-offs are most visible in the \textit{lit\_review} and \textit{development} labels. The classifier suffers from very low recall for \textit{lit\_review} ($0.09$), as these sections are frequently titled with descriptive, discipline-specific headers that do not match generic dictionary terms. However, the \textit{development} label shows a high recall ($0.79$) but extremely low precision ($0.12$). This confirms that while our rule-based detection of numbered top-level headers successfully flags these argumentative sections, it often over-captures them, reflecting the difficulty of distinguishing "development" from the surrounding narrative without full-text semantic analysis. Crucially, identifying these sections serves a vital structural purpose by helping to define the terminal boundary of the introduction, improving the overall accuracy of the major IMRaD labels.

\subsection{Error Analysis and Confusion Matrix}

\begin{figure}[h!]
    \centering
    \includegraphics[width=0.85\textwidth]{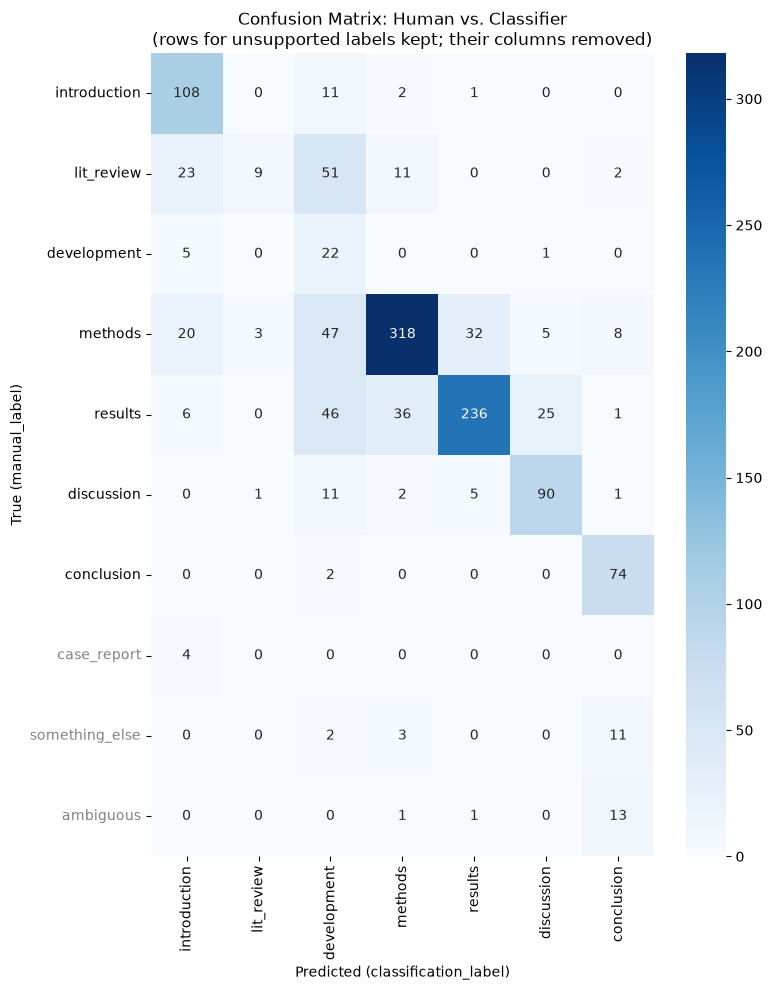} 
    \caption{Confusion Matrix: Rule-Based Classifier (Predictions) vs. Human Gold Standard (True Labels).}
    \label{fig:confusion_matrix}
\end{figure}

To visualize the specific misclassifications driving the metrics discussed above, the confusion matrix in Figure \ref{fig:confusion_matrix} compares the rule-based classifier against the human gold standard. The matrix ``diagonal" (where the manual and classification labels match) shows strong performance for the core empirical sections. However, the non-diagonal densities reveal the specific nature of the challenges at the fine-grained level.

The matrix highlights significant discrepancies in the attribution of the \textit{lit\_review} and \textit{development} labels. For \textit{lit\_review}, the classifier was extremely conservative, assigning the label to very few cases overall. Notably, 23 sections that the classifier labeled as \textit{introduction} were identified by human annotators as \textit{lit\_review}, confirming that the classifier's broad definition of the introduction often subsumes these more specific rhetorical sections. 

%%The \textit{development} label, on the other hand, shows widespread confusion. When the classifier predicts \textit{development}, the human labels are distributed across several core categories: 51 instances were labeled as \textit{lit\_review}, 47 as \textit{methods}, and 46 as \textit{results}. This confirms that using top-level numbered headers as a proxy for argumentative "development" is a noisy heuristic that catches a wide variety of content.

The \textit{development} label exhibits the greatest ambiguity among all categories. Unlike the core IMRaD labels, it is defined structurally rather than rhetorically, capturing major numbered sections that are not matched by existing section dictionaries. Consequently, the label overlaps with several substantive section types. In the manual annotations, sections classified as \textit{development} were frequently labeled as \textit{lit review}, \textit{methods}, or \textit{results} by human annotators. The larger LLM evaluation shows a similar pattern, with substantial confusion between \textit{development} and multiple standard section categories. These results suggest that the \textit{development} label often captures sections that serve recognizable rhetorical functions but use discipline-specific or otherwise uncommon titles that are not covered by our rule-based patterns. At the same time, the label remains useful for identifying major section boundaries, particularly the transition from introductory material into the main body of the paper. Future work could potentially refine these sections using large language models, which may be better able to infer rhetorical function from section content rather than relying primarily on section titles and document structure.

Finally, we observe expected rhetorical overlaps between adjacent sections, such as the 36 instances where humans labeled sections as \textit{results} but the classifier identified them as \textit{methods}, and the overlap between \textit{results} and \textit{discussion} (25 instances). 

Overall, the validation results demonstrate that the rule-based classifier provides reliable section-level annotations suitable for large-scale analysis, while acknowledging inherent ambiguity in some classification cases that affects all annotation approaches.

\section{Usage Notes}

Researchers using this dataset should be aware of several characteristics and limitations. The dataset focuses on empirical papers with IMRaD-like structures, so it has better coverage of STEM disciplines (approximately 75 percent of papers) than humanities and social sciences. Within the dataset, medicine is highly represented (25 percent of all papers).

Most sections in the corpus correspond to a single section label. However, some papers use headers that combine multiple functions, such as ``Results and Discussion" or ``Discussion and Conclusion". To accommodate these cases, we provide an additional column containing alternative multi-category labels. These labels are relatively uncommon overall. The most frequent case is ``Results and Discussion", which occurs for approximately 18\% of sections assigned the discussion label, while other combined categories occur in less than 3\% of their respective section types. We therefore use a single primary label throughout the main annotation scheme to maintain a consistent section taxonomy, while preserving combined labels in a supplementary field for researchers interested in finer-grained distinctions.

Approximately 8 to 10 percent of papers in S2ORC contain substantial content in languages other than English, despite the corpus's intended language restriction. Users conducting English-language text analysis should consider filtering by language or implementing language detection.

The section labels reflect a standardized schema that may not capture all discipline-specific variations in section naming or organization. The development label indicates major sections after the introduction that could not be classified into standard categories, often representing argumentation sections, literature reviews, or other discipline-specific organizational patterns.

Label propagation assumes that all unlabeled headers between two labeled headers belong to the preceding section. This assumption works well for papers with clear major section boundaries but may misclassify subsection headers or nested structures. Users should consider this when analyzing section-internal organization or working with papers that have complex hierarchical structures.

To maximize utility and flexibility for the research community, we release the full, unfiltered, and unaggregated dataset of section labels. We do not distribute the aggregated dataset containing the parsed text due to significant file size constraints, and because users must retrieve the underlying text by joining our provided document identifiers directly with the Semantic Scholar Open Research Corpus (S2ORC) \footnote{\url{https://github.com/allenai/s2orc}}. Furthermore, providing the raw classification outputs allows users to apply custom quality thresholds. Researchers can easily implement their own filtering logic (such as adjusting the minimum section counts or maximum length proportions) tailored to the specific needs of their downstream applications. Our filtering criteria seek to retain well-structured empirical papers and may exclude some argumentative essays, reviews, or papers with non-standard structures. Therefore, researchers interested in these document types may need alternative approaches.

\section{Data Availability}

The complete section-level classification dataset generated in this study is publicly available in Parquet format via Zenodo at \url{https://doi.org/10.5281/zenodo.20814302}. This repository contains the complete mapping of S2ORC paper identifiers, section labels, and document position metadata, along with the code required to reproduce the classification pipeline and documentation of the schema. 

Because our dataset links directly to the original S2ORC corpus to avoid textual redundancy, users can fully reconstruct the text-label pairs by matching our paper identifiers to the publicly available S2ORC dataset.

\section{Code Availability}

The complete classification pipeline, including data preparation, labeling algorithms, label refinement, and filtering steps, is available at \url{https://github.com/daniel-verdi/section-splitter}. The code is written in Python and uses standard scientific computing libraries, including pandas for data manipulation and regular expression libraries for pattern matching. Documentation includes instructions for reproducing the classification process and applying the pipeline to new data.

\section{Acknowledgements}

We thank all human annotators who contributed to the validation effort. We acknowledge the Semantic Scholar team for providing the S2ORC dataset. We are also grateful to the members of the Sinatra Lab, the Networks, Data, and Society (NERDS) research group, and collaborators at the University of Copenhagen and at the IT University of Copenhagen for valuable discussions and feedback throughout this project.

\section{Author Contributions}
J.A.D. conceived the project. D.V. developed the classification algorithm, conducted the analysis, performed validation studies, and drafted the manuscript. J.A.D. provided mentorship and guidance throughout the project and set up the annotation study on Prolific. R.S. supervised the research and provided critical feedback on methodology and presentation. All authors reviewed and approved the final manuscript.

\section{Competing Interests}

The authors declare no competing interests.

\section{Funding}

All authors acknowledge support from Villum Fonden through the Villum Young Investigator programme (project number: 00037394). D.V. also acknowledges support from the Danish Data Science Academy (DDSA) through the Visit Grant. This work is co-funded by the European Research Council (ERC) grant \textit{scAIence}, No. 101125480. Co-funded by the  European Union. Views and opinions expressed are those of the authors only and do not necessarily reflect those of the European Union or the European Research Council. Neither the European Union nor the granting authority can be held responsible for them.

%\bibliographystyle{unsrt}
%\bibliography{refs}

\begin{thebibliography}{10}

\bibitem{imrad2004}
Luciana~B. Sollaci and Mauricio~G. Pereira.
\newblock The introduction, methods, results, and discussion (imrad) structure: A fifty-year survey.
\newblock {\em Journal of the Medical Library Association}, 92(3):364, 2004.

\bibitem{imrad2024}
Cary Moskovitz, Benjamin Harmon, and S.~Saha.
\newblock The structure of scientific writing: An empirical analysis of recent research articles in stem.
\newblock {\em Journal of Technical Writing and Communication}, 54(3):265--281, 2024.

\bibitem{shahid2018section}
Abdul Shahid and Muhammad~Tanvir Afzal.
\newblock Section-wise indexing and retrieval of research articles.
\newblock {\em Cluster Computing}, 21(1):481--492, 2018.

\bibitem{treeratpituk2010seerlab}
Pucktada Treeratpituk, Pradeep Teregowda, Jian Huang, and C~Lee Giles.
\newblock Seerlab: A system for extracting keyphrases from scholarly documents.
\newblock In {\em Proceedings of the 5th International Workshop on Semantic Evaluation}, pages 182--185, 2010.

\bibitem{tuarob2015hybrid}
Suppawong Tuarob, Prasenjit Mitra, and C~Lee Giles.
\newblock A hybrid approach to discover semantic hierarchical sections in scholarly documents.
\newblock In {\em 2015 13th international conference on document analysis and recognition (ICDAR)}, pages 1081--1085. IEEE, 2015.

\bibitem{nguyen2007}
Thuy~Dung Nguyen and Min-Yen Kan.
\newblock Keyphrase extraction in scientific publications.
\newblock In Dion Hoe-Lian Goh, Thanh~Hoang Cao, Ingeborg~T. S{\o}lvberg, and Edie Rasmussen, editors, {\em Asian Digital Libraries. Looking Back 10 Years and Forging New Frontiers}, pages 317--326. Springer, 2007.

\bibitem{rahman2019_unfolding}
Muhammad~Mahbubur Rahman and Tim Finin.
\newblock Unfolding the structure of a document using deep learning, 2019.

\bibitem{lo2020s2orc}
Kyle Lo, Lucy~Lu Wang, Mark Neumann, Rodney Kinney, and Daniel~S. Weld.
\newblock S2orc: The semantic scholar open research corpus.
\newblock In {\em Proceedings of the 58th Annual Meeting of the Association for Computational Linguistics}, pages 4969--4983, 2020.

\bibitem{fortunato2018science}
Santo Fortunato, Carl~T Bergstrom, Katy B{\"o}rner, James~A Evans, Dirk Helbing, Sta{\v{s}}a Milojevi{\'c}, Alexander~M Petersen, Filippo Radicchi, Roberta Sinatra, Brian Uzzi, et~al.
\newblock Science of science.
\newblock {\em Science}, 359(6379):eaao0185, 2018.

\bibitem{grobid}
Grobid.
\newblock \url{https://github.com/grobidOrg/grobid}, 2008--2025.

\bibitem{s2_fos}
{Allen Institute for AI}.
\newblock S2 fos.
\newblock \url{https://github.com/allenai/s2_fos}.
\newblock Computer software. Accessed June 2026.

\bibitem{core}
Petr Knoth, Drahomira Herrmannova, Matteo Cancellieri, Lucas Anastasiou, Nancy Pontika, Samuel Pearce, Bikash Gyawali, and David Pride.
\newblock Core: A global aggregation service for open access papers.
\newblock {\em Scientific Data}, 10(1):366, 2023.

\bibitem{openalex}
{OpenAlex Developers}.
\newblock Full-text pdfs: Download full-text pdfs and tei xml for millions of works.
\newblock \url{https://developers.openalex.org/download/full-text-pdfs}.
\newblock OpenAlex Developer Documentation. Accessed June 2026.

\end{thebibliography}

\end{document}